\documentclass[preprint,11pt, nofootinbib]{revtex4-1}

\usepackage{amssymb,amsmath,setspace,graphicx,mathrsfs, float,slashed, mathrsfs, amsfonts, setspace, natbib, gensymb, siunitx}

\usepackage[usenames, dvipsnames]{color}

\usepackage[bottom]{footmisc}

\begin{document}
	
\title{Signature of light sterile neutrinos at IceCube}

\date{\today}

\author{Bhavesh Chauhan}
\email{bhavesh@prl.res.in}
\affiliation{Physical Research Laboratory, Ahmedabad, India}
\affiliation{Indian Institute of Technology, Gandhinagar, India}

\author{Subhendra Mohanty}
\email{mohanty@prl.res.in}
\affiliation{Physical Research Laboratory, Ahmedabad, India.}

\begin{abstract}
The MiniBooNE collaboration has recently reported evidence for a light sterile neutrino with large mixing angles  thus corroborating the measurement by LSND twenty years ago. Such a state would be directly in conflict with Planck measurement of BBN $N_{eff}$ unless there is self-interaction in the sterile sector. Our objective is to investigate if such interactions could result in resonant absorption in the cosmogenic neutrino spectrum and its consequences for the IceCube experiment. We show that it is possible to give independent bounds on sterile neutrino parameter space from IceCube observations with the dips in the spectrum corresponding to the neutrino masses.
	
\end{abstract}

\maketitle

\section{Introduction}

The MiniBooNE collaboration has recently reported excess in the electron neutrino and antineutrino appearance channels that is consistent with the sterile neutrino hypothesis \cite{miniboone2018}. The best-fit point,
\begin{equation}
\Delta m_{41}^2 = 0.041~eV^2  \quad and \quad \sin^2(2 \theta_{\mu e}) = 0.958
\end{equation}
is consistent with the earlier measurements by the LSND collaboration \cite{lsnd}. In fact, the combined significance of the two data sets is 6.1$\sigma$. These results, however, are in tension with data from disappearance experiments like MINOS+ and IceCube. Other experiments like KARMEN and OPERA have not been able to confirm this excess, but they do not rule it out completely either  \cite{updatedfits}.

The existence of such a light states with large mixing angles is also in conflict with cosmology. The Planck measurement Cosmic Microwave Background (CMB) anisotropy puts severe constraints on number of thermalised relativistic degrees of freedom ($N_{eff}$) around the epoch of Big-Bang Nucleosynthesis (BBN) i.e. $T_\gamma = 1$ MeV \cite{planck} . One possible resolution to this puzzle is to assume self-interactions in the sterile sector \cite{Hannestad2013, Basudeb2013, Basudeb2015, Archidiacono1, Archidiacono2, Archidiacono3}. Due to the large thermal effective potential, the mixing between sterile and active neutrino is suppressed in the early universe but is allowed to be large today. Hence the sterile neutrinos are produced efficiently only at low temperatures after recoupling \cite{Saviano}. This provides a very strong constraint that the $T_{rec} < 1$ MeV which rules out small gauge couplings in the sterile sector \cite{Basudeb2015}.  Due to mixing, the lighter neutrinos also interact with the new gauge boson which affects their free streaming in the early universe which is constrained from CMB \cite{MirizziSumNeu, Forastieri:2017oma}.  It was recently pointed out that taking constraints from $\sum m_\nu$ rules out any viable parameter space for $m_s > 0.2 $ eV \cite{Basudeb2018}. However, the authors also propose several scenarios which weaken these new constraints. For gauge coupling in the range 0.1 - 1 , one requires a gauge boson of mass 10 - 50 $MeV$ to reconcile sterile neutrinos with cosmology. Moreover, such interactions can also be mediators to dark matter which can simultaneously solve the small-scale crisis of $\Lambda CDM$ \cite{Chauhan, Bringmann, Basudeb2013}. \\  

It was shown in \cite{IbeKaneta} that MeV scale secret interaction of neutrinos will give rise to absorption lines in the very high energy neutrino spectrum. Such lines can be seen by neutrino telescopes like IceCube. The IceCube HESE data has featured a prominent gap in the spectrum for neutrino energies in the range 400-800 TeV \cite{IceCube3, IceCube4, IceCube6}. In the past, several authors have tried to explain this gap using resonant absorption in well motivated models such as $\nu$2HDM \cite{IbeKaneta} and gauged $U(1)_{L_\mu - L_\tau}$ \cite{Araki}. Recently it was also proposed that one can explain the absence of Glashow resonance using t-channel resonant absorption \cite{Ashish}. All these explanations assume a flavor-universal single power law flux for incoming neutrinos. The IceCube data can also be explained by decaying dark matter \cite{Atri1, Atri2, Dhuria, Rott:2014kfa, Dev:2016qbd, Sui:2018bbh}, Leptoquark like states \cite{Anchordoqui:2006wc, Barger:2013pla, Dev:2016uxj, Mileo:2016zeo, Dey:2015eaa}, and my modifying assumptions of the source. The Leptoquark explanation is highly constrained from LHC data \cite{Mileo:2016zeo, Chauhan:2017ndd, Dey:2017ede}. \\

In this paper, we look at resonant absorption of cosmogenic neutrinos from both cosmic neutrino and sterile neutrino background. In section II we describe the model for sterile neutrino with self interactions. In Section III, we discuss the basics of neutrino absorption and explain a few benchmark scenarios. In Section IV we look at the six year IceCube data and provide some constraints on the model. We also provide the parameter space favored by IceCube independent of other short baseline experiments. In section V we provide the results and discuss certain aspects of the analysis before we conclude.

\section{Model Description and Cosmological Constraints}

To accommodate light sterile neutrino with cosmology, we extend the Standard Model by introducing a left-handed sterile neutrino ($\nu_s$) which is charged under an additional gauge symmetry $U(1)_X$. The new gauge boson ($X_\mu$) would acquire its mass through spontaneous symmetry breaking in the hidden sector. The scalar responsible for the phase transition can also thermalise the sterile sector in early universe through Higgs' portal. The requirement of anomaly cancellation needs additional fermions in the spectrum which can be a dark matter candidate. However, for our analysis, we only focus on the sterile neutrino and its interactions. \\

The relevant part of the Lagrangian is the gauge interaction of the sterile neutrino which is given by, 
\begin{equation}
\label{sl}
- \mathcal{L}_s = g_X \bar{\nu}_s \gamma^\mu P_L \nu_s X_\mu
\end{equation}
In terms of mass eigenstates, 
\begin{equation}
- \mathcal{L}_s = \sum_{i,js} g_{ij} \bar{\nu}_i \gamma^\mu P_L \nu_j X_\mu
\end{equation}
where  $g_{ij} = g_X U^{\ast}_{s i} U_{ s j}$. The $4 \times 4$ PMNS matrix is parametrised as, 
\begin{equation}
U = R_{34} R_{24} R_{14} R_{23} R_{13} R_{12}
\end{equation}
where $R_{ij}$ is the rotation matrix in the i-j plane. We assume that the elements of the mixing matrix are real as contribution of the phases is negligible for the discussion that follows. We also fix the active neutrino mixing angles to the best-fit values from the oscillation measurements \cite{Esteban:2016qun}, 
\begin{equation}
\theta_{12} = \ang{33.62} \quad \theta_{23} = \ang{47.2} \quad \theta_{13} = \ang{8.54} 
\end{equation}
We have 6 free parameters in our model, 
\begin{equation}
\mathcal{P} = \{ \theta_{14}, \theta_{24}, \theta_{34}, m_4, g_X, M_X \}.
\end{equation}
where $m_4$ is the mass of the fourth (mostly sterile) mass eigenstate and $M_X$ is the mass of new gauge boson. \\

The introduction of self interactions generates a finite temperature effective potential for the sterile neutrino of the form \cite{Basudeb2013}, 
\begin{equation}
V_{eff}=\begin{cases}
- \frac{28 \pi^3 \alpha_X E T_s^4}{45 M_X^4} \quad ~E, T_s \ll M \\
+ \frac{\pi \alpha_X T_s^2}{2 E}  \quad \quad \quad E, T_s \gg M
\end{cases}
\end{equation}
which modifies the effective mixing angle given by, 
\begin{equation}
\sin^2(2 \theta_m) = \frac{\sin^2 (2 \theta_0)}{(\cos(2\theta_0) + \frac{2E}{\Delta m^2} V_{eff})^2 + \sin^2 (2 \theta_0)}.
\end{equation}
In the early universe when the temperature is high, the mixing angle is suppressed and the production rate of the sterile neutrino is negligible. As the universe cools, the sterile sector recouples to the Standard Model bath. If the recoupling temperature is $>$ MeV, then the sterile neutrinos are thermalised before the Big Bang Nucleosynthesis takes place. Since they are relativistic during BBN, there are very stringent constraints from Planck. Hence, one requires the recoupling temperature to be less than an MeV. In \cite{Basudeb2018} it was shown that the entire parameter space for the scenario is ruled out for $m_4 \geq 1~eV$. However, it was also pointed out that there are several possible new physics effects that can alleviate these bound. One of the plausible scenario is where one adds new lighter particles in the model. 

\section{Neutrino absorption by Cosmic Neutrino Background}

Until very recently, the source of ultra-high energy neutrinos was unknown. Advances in multi-messenger astronomy have pointed towards blazars as possible sources \cite{IceCube:2018dnn}. During propagation through the cosmic media, these neutrinos can get resonantly scattered off the cosmic neutrino background which results in an absorption line in the neutrino spectrum. If only Standard Model interactions are considered, the absorption line ($\sim 10^{13}$ GeV) is undetectable at neutrino telescopes \cite{Weiler:1982qy}. However, it has been known that secret interaction of the neutrino can also give rise to these lines which should, in principle, be detectable \cite{Ng:2014pca, Ioka:2014kca, Blum:2014ewa}. The absorption lines from sterile neutrino were first pointed out in \cite{Cherry:2016jol}, and \cite{Jeong:2018yts} applied it in the context of diffuse supernova background. In this paper, we attempt to explain the two dips in the IceCube spectrum using resonant absorption by heavy mostly sterile and the heaviest active neutrino.\\

We have assumed that, due to recoupling of the sterile neutrinos, the neutrino background has all four mass eigenstates in equal proportions and at the same temperature. For the benchmark scenarios considered in the paper, the recoupling is guaranteed \cite{Basudeb2015}. The scattering cross section is, 
\begin{equation}
\sigma_{ij} = \sigma \left( \bar{\nu}_i \nu_j \rightarrow \bar{\nu} \nu \right) = \frac{1}{6 \pi} |g_{ij}|^2 g_{X}^2 \frac{s}{(s - m_{X}^2)^2 + m_{X}^2 \Gamma_{X}^2}
\end{equation}
where $\nu_i$ are the mass eigenstates of the four neutrino species and $\Gamma_{X} = g_X^2 m_X / 12 \pi $ is the decay width of the new boson. The mean free path is, 
\begin{equation}
\label{mfp}
\lambda_i (E_i, z) =  \left( \sum_{j} \int \frac{d^3 \mathbf{p}}{(2 \pi)^3} f_{j} ( p ,z) \sigma_{ij}( p, E_i, z) \right)^{-1} \approx  \left( n_\nu(z) \sum_{j} \sigma_{ij}( p, E_i, z) \right)^{-1}
\end{equation}
where $f_i$ is the distribution function for the neutrinos given by, 
\begin{equation}
f_i (p, z) ^{-1} = \exp \left( \frac{p}{T_{i}(1+z)} \right) + 1 
\end{equation}
and $T_{i} = 1.95~K$ for all four components. The approximation in the RHS of Eq. \eqref{mfp} is valid only when the neutrino is non-relativistic. The oscillation data suggests that at least two active neutrinos are non-relativistic today. As we shall see, the lightest neutrino gives absorption feature for higher energies and is inconsequential to our discussion. For the remainder of the paper, we assume Normal Hierarchy and neutrino masses to be 
\begin{equation}
m_1 = 5\times 10^{-3}~eV,\quad m_2= 1 \times 10^{-2}~eV,\quad m_3 = 5 \times 10^{-2}~eV.
\end{equation}
The case of Inverted Hierarchy is commented upon at the end of this section. One can see that,
\begin{equation}
m_i \gg \langle p \rangle = 3 T_\nu \sim 5.3 \times 10^{-4}~eV ~ \forall~i. 
\end{equation}
which allows us to approximate, 
\begin{equation}
s = 2 E_i (1 + z) \left( \sqrt{p^2 + m_i^2} - p \cos[\theta] \right) \approx 2 E_i (1+z) m_i.
\end{equation}
The $z$ dependence accounts for redshift during propagation. The survival rate of neutrino is given as \cite{Blum:2014ewa,DiFranzo:2015qea},
\begin{equation}
R_i = \exp \left[   - \int_{0}^{z_s} \frac{1}{\lambda_i (1 + z )} \frac{dL}{dz} dz\right]
\end{equation}
where $z_s$ denotes the redshift distance to the source and,
\begin{equation}
\frac{dL}{dz} = \frac{c}{H_0 \sqrt{ \Omega_m (1 + z)^3 + \Omega_\Lambda}}. 
\end{equation}
We have fixed the cosmological parameters to $\Omega_m = 0.315, ~\Omega_\Lambda = 0.685, ~H_0 = 67.3$ km/s/Mpc using the best fit values from Planck \cite{planck}. We also assume a power-law flux for each neutrino near the source. The flux of neutrino of flavor $\alpha \in {e, \mu, \tau, s}$ at Earth is, 
\begin{equation}
\phi_\alpha = \sum_{j=1}^{4} |U_{\alpha j}|^2 \phi_j R_j  =  (\phi_0 E_\nu^{-\gamma}) \sum_{j=1}^{4} |U_{ \alpha j}|^2 R_j   \equiv  (\phi_0 E_\nu^{-\gamma}) R_\alpha.
\end{equation}
Since the sterile neutrino will not generate any signal at the IceCube detector, the flux of neutrinos that can be seen by IceCube is simply, 
\begin{equation}
\phi = \phi_e + \phi_{\mu} + \phi_\tau = (\phi_0 E_\nu^{-\gamma}) \left( \sum_{f= e,\mu,\tau} \sum_{j=1}^{4} |U_{fj}|^2 R_j  \right) \equiv \phi_0 E_\nu^{-\gamma} \langle R  ( \mathcal{P}, E_\nu)\rangle
\end{equation}
where the parentheses in the last part indicate that $\langle R \rangle$ depends on the model parameters and incident neutrino energy only. \\

In Fig (1), we have shown the variation of $R_\alpha$ and $R_i$ for a benchmark scenario. The gauge coupling is fixed to be $g_X = 0.1$ and the mass of the gauge boson to be $M_X = 25$ MeV. We have assumed that the neutrino sources are localised around $z_s$ = 0.3. There are three features we would like to highlight: (a) There are two prominent dips in the function. The one at lower neutrino energy is associated with the absorption due to heavy (i.e. mostly sterile) mass eigenstate. The second dip is due to the absorption by the heaviest active neutrino ( i.e. $m_3$ in NH). (b) The dips are not very sharp and there is a broadening due to redshift during propagation. For a source located at $z_s$, the dip in the spectrum occurs for the neutrino energies
\begin{equation}
E_{dip} : \frac{E^{res}}{(1 + z_s)} \rightarrow E^{res}
\end{equation}
where $E^{res} = M_X^2/2 m_i$. This allows us to estimate the width of the dip as, 
\begin{equation}
\label{width}
\Delta^i \approx \frac{M_X^2}{2 m_i} \frac{z_s}{1 + z_s} .
\end{equation}(c) Since the other active neutrinos are lighter, their absorption lines are at much higher neutrino energies. Hence, it is inconsequential for our analysis whether the lightest neutrino is relativistic or non-relativistic today. \\

The absorption lines are sensitive to the distance to the source. It can be inferred from \eqref{width} that the further the source, the broader will be the absorption line. We have assumed that the UHE neutrinos originate from blazars and non-blazar AGNs as opposed to spatially distributed sources like Dark Matter decay \cite{IceCube:2018dnn, Krauss:2014tna, Sahu:2017spx, Miranda:2015ema, Hooper:2018wyk}. Future multi messenger observations will help us verify this hypothesis. For this analysis we assume that the sources are localised around a particular redshift, $\langle z_s \rangle$, which makes the calculations simple. The complete analysis which also considers distribution of the sources is beyond the scope of this work. Also note that, any source located very far from Earth ($z_s > 5$) will have too broad absorption lines and contribute negligibly to the flux at high energies ($>$ 200 TeV). This may be compatible with the fact that IceCube rarely sees events of such high energies. This inference cannot be made in the standard picture without secret interactions. Thus, if future multi-messenger observations infer that almost all the sources of UHE neutrinos are localized within a sphere, it will strongly hint at resonant absorption. \\

\begin{figure}
	\centering
	\begin{tabular}{c c}
		\includegraphics[width=8cm]{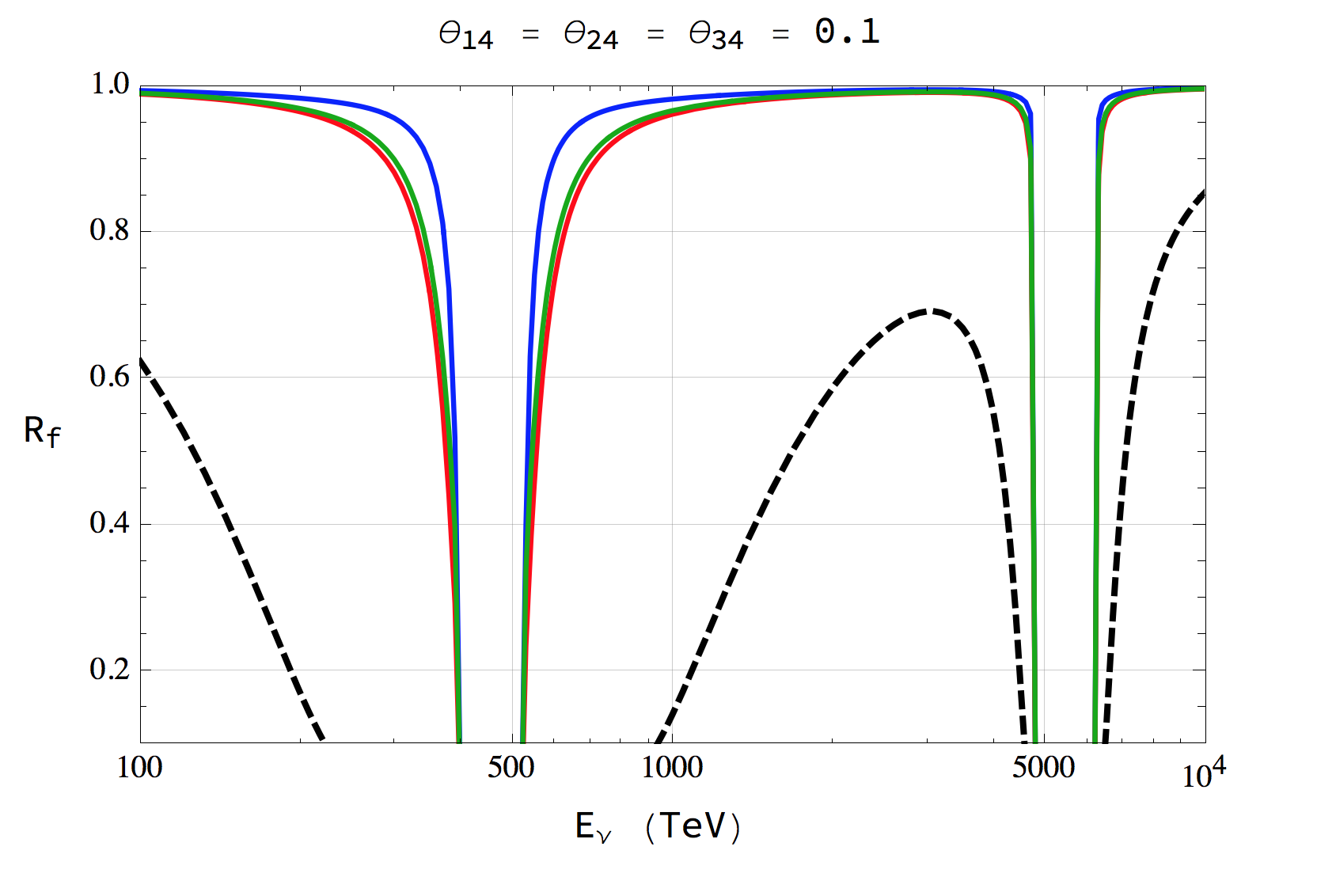}  & 	\includegraphics[width=8cm]{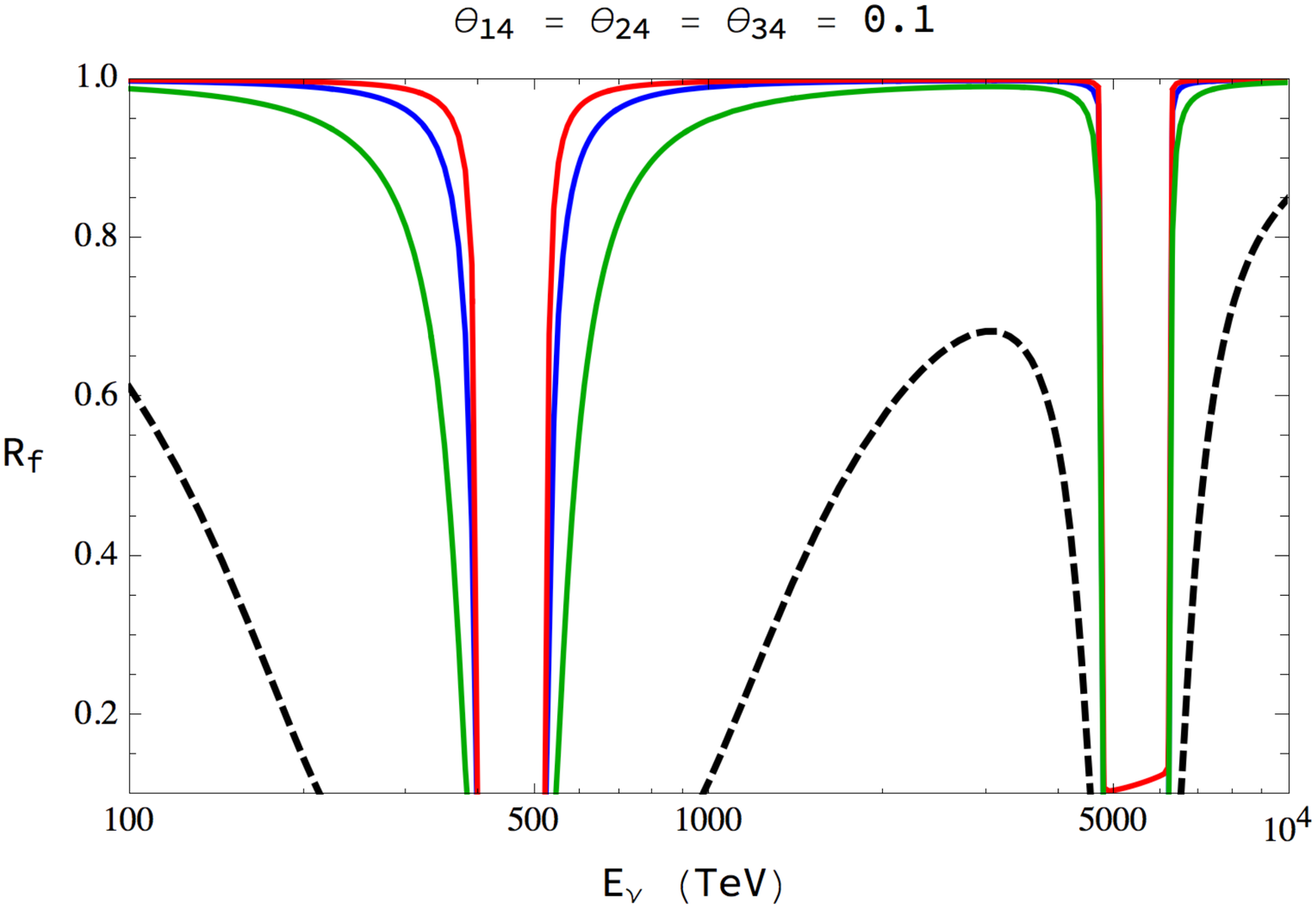}
	\end{tabular}
	\caption{ \emph{Left:} This plot shows variation of $R_e$ (Blue), $R_\mu$ (Red), $R_\tau$(Green), and $R_s$(Black, Dashed) with neutrino energy. \emph{Right:} This plot shows variation of $R_1$ (Blue), $R_2$ (Red), $R_3$(Green), and $R_4$(Black, Dashed) with neutrino energy. See text for details.}
\end{figure}

\section{Constraints from flux of neutrinos at IceCube}

In IceCube six-year HESE data, 82 events passed the selection criterion of which two are co-incident with atmospheric muons and left out \cite{IceCube6}. The best fit for single power law flux is, 
\begin{equation}
E_\nu^2 \phi = (2.46 \pm 0.8) \times 10^{-8} 
\left( \frac{E_\nu}{100~TeV}
\right)^{-0.92} 
 ~ GeV cm^{-2}s^{-1}sr^{-1}
\end{equation}
which has softer spectral index than the 3-year ($\gamma = 2.3 $) \cite{IceCube3} as well as the 4-year ($\gamma = 2.58 $) data \cite{IceCube4}. Once can attribute this to the pile-up of low energy events along with the lack of high energy events in the new data. A prominent feature that still remains is the apparent lack of neutrinos with energy 400-800 TeV. From one point of view, one should be able to see these neutrinos with more exposure. However, this may also hint at new physics. Another puzzling mystery is the absence of Glashow Resonance. In the Standard Model, the astrophysical neutrino can interact with the electrons in the detector volume and produce an on-shell W-boson. This happens for neutrino energy $\sim$ 6.3 PeV. Around this energy, the cross section for neutrino-electron scattering is several orders of magnitude larger than the charged and neutral current interactions with nucleons. Thus we expect more number of events in the 3.6 PeV to 7.5 PeV bin. Due to this, the best fits to the data hint towards a softer spectral index. Several scenarios have been proposed to address the absence of Glashow events including active neutrino decay, $\Delta^+$ resonance, and novel flux \cite{Pakvasa:2012db, Sahu:2016qet, Kistler:2016ask}.\\

Now we examine the $m_s - M_X$ parameter space that can explain the observed IceCube spectrum. The following constraints are imposed:
\begin{enumerate}
	\item If $E^{res} \sim PeV$, one cannot explain the observed PeV events at IceCube unless exceptional circumstances are evoked. To be general, we constrain the $m_3$ absorption line to be more than 3 PeV. Because of the broadening during propagation, the constraint depends on $\langle z_s \rangle$ as 
	\begin{equation}
	M_X^2 \geq 2 \times 3~ PeV~m_3~(1 + \langle z_s \rangle)
	\end{equation}
	This is shown in Fig. (2) as region bounded by green lines. 
	
	\item Since we wish to explain the dip in the spectrum using the fourth neutrino, we require,
	\begin{equation}
	E^{res} \leq~800~TeV \quad \text{\&} \quad \frac{E^{res}}{1 +\langle z_s \rangle } \geq~400~TeV
	\end{equation}
	which is shown as the blue shaded region in Fig. (2). 
	
	\item We show the region in the parameter space that requires more than 1, 2, and 3 lighter sterile neutrinos in the full theory (cf. Eq. (10)). 
\end{enumerate}

It can be seen from Fig. (2) that only a small portion of the parameter space is compatible with all the constraints. With slightly relaxed assumptions, we chose the representative point 
\begin{equation}
m_4~=~0.4~eV \quad \text{\&} \quad M_X~=~25~MeV
\end{equation} 
for our analysis. The gauge coupling is constrained from the restrictions on the recoupling temperature. We have chosen the benchmark point $g_X = 0.1$ which is consistent. \\

\begin{figure}
	\centering
	\includegraphics[width=11cm]{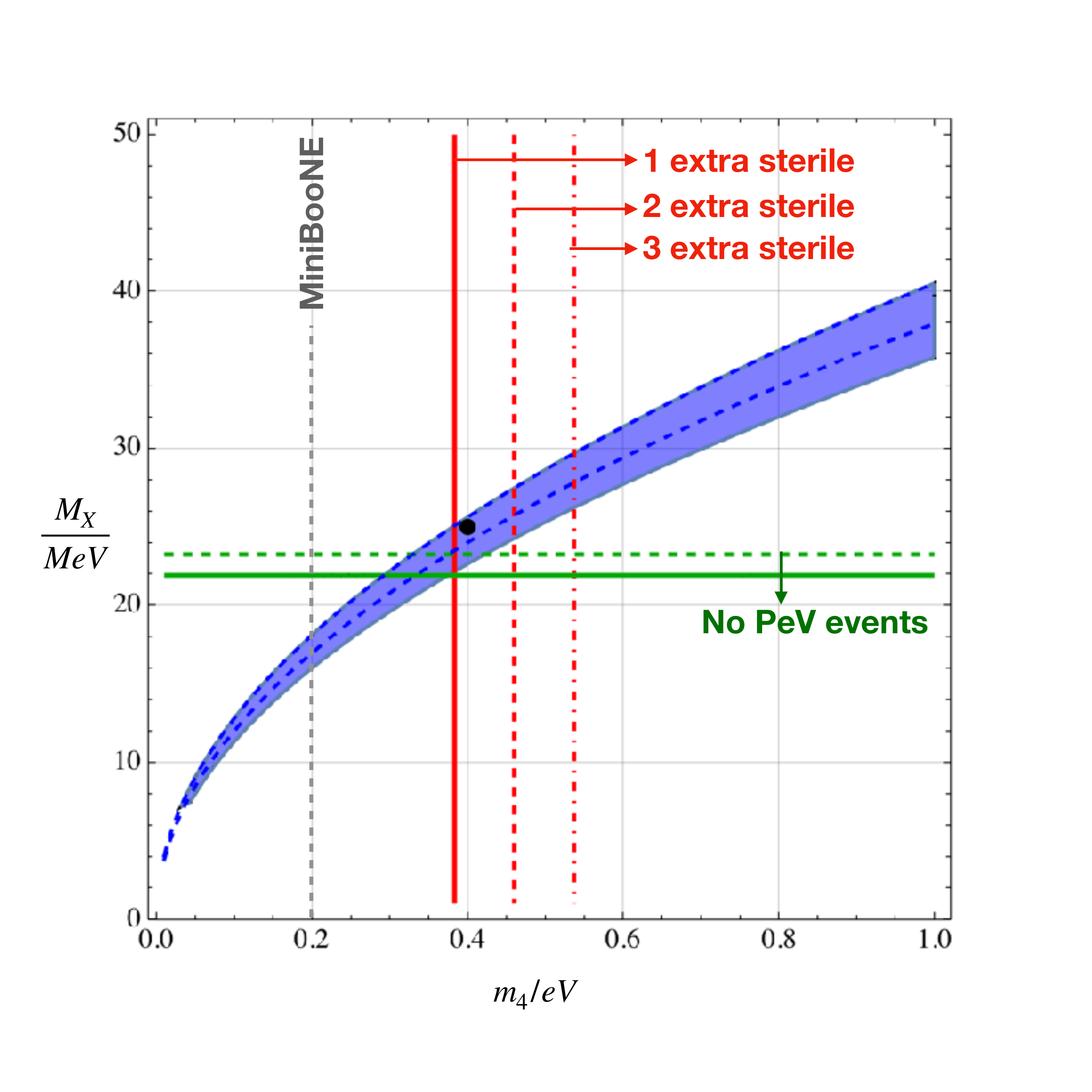}
	\caption{The shaded blue region with solid (dashed) boundaries can explain the 400-800 TeV dip in the IceCube spectrum assuming that the source are distributed around z = 0.6 (0.8). The solid (dashed) green lines denote the upper bound on X boson mass such that the gap due to heaviest active neutrino is above 3 PeV assuming source distribution around z = 0.6 (0.8). The green arrows indicate the region that is disfavored. The red lines (solid, dashed, dot-dashed) denote the number of additional light particles (1,2,3) to be added to the theory to evade $\sum m_\nu$ constraints. The black point shows the benchmark case considered in the paper. The MiniBooNE best-fit is highlighted. See text for more details. }
\end{figure}

\begin{figure}
	\centering
	\includegraphics[width=11cm]{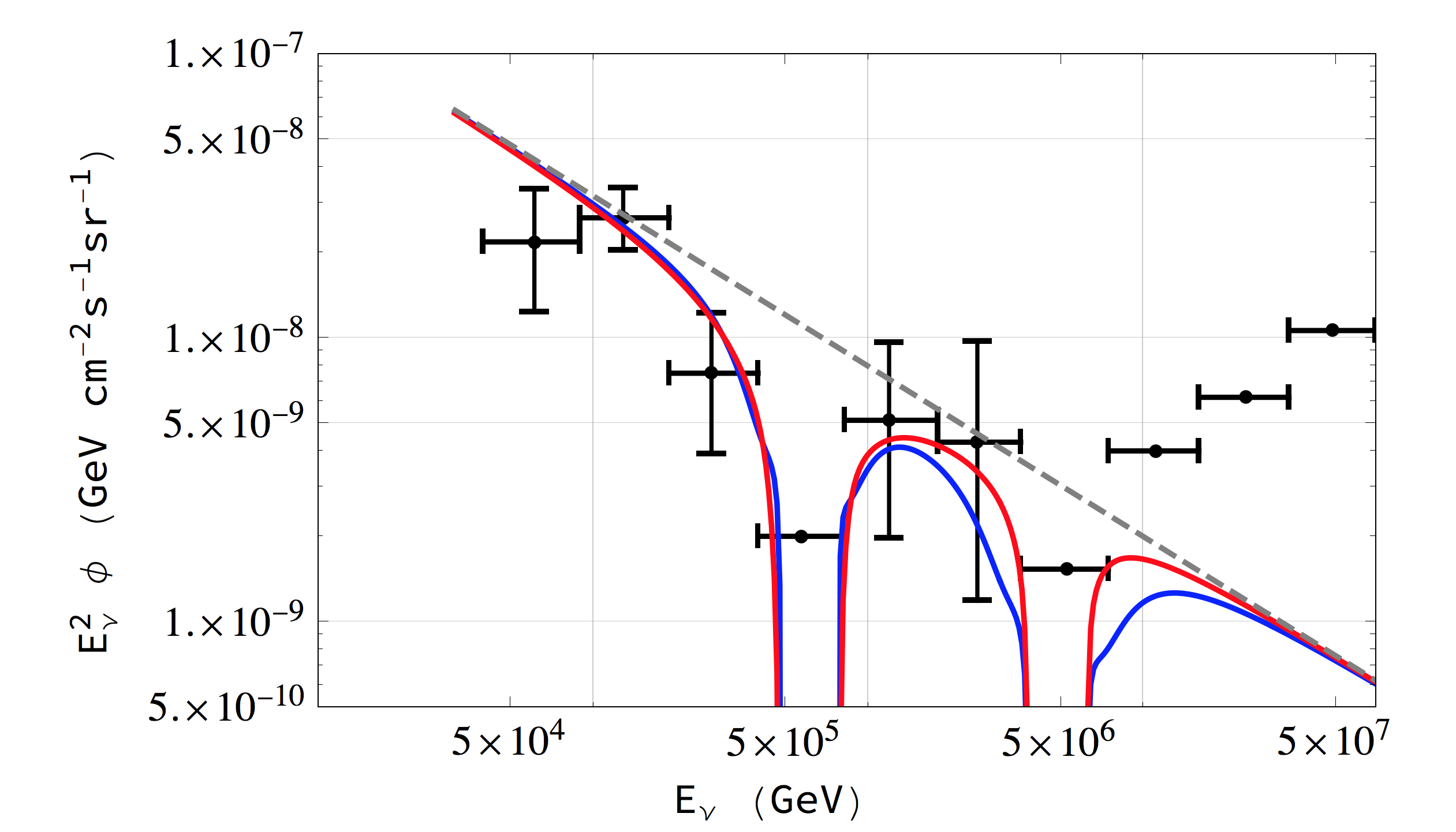}
	\caption{ \label{res} The flux without attenuation is shown as dashed gray curve. The blue (red) curve is the flux with attenuation for the democratic (maximal) case. The spectral index is chosen to be 2.6 and the normalization is fixed from the second bin. Sources are assumed to be distributed around z = 0.6.  }
\end{figure}

For the choice of mixing angles, we have considered two scenarios, 
\begin{align}
\text{Case I} &: \quad \theta_{14} = \theta_{24} = \theta_{34} = 0.3 \quad\quad \quad\quad\quad\quad \quad ~ ~ ...(democratic)\\
\text{Case II} &: \quad \theta_{14} = \theta_{24} = \pi/4 \quad \text{\&}\quad \theta_{34} = 0 \quad \quad\quad\quad...(maximal).
\end{align}
For the democratic case, we have checked that the choice 0.3 gives the best fit to the data. The maximal case is motivated by the mixing angles observed by MiniBooNE. We have chosen the spectral index to be 2.6 which is consistent with IceCube best fits. Any softer spectral index will result in reducing the flux of PeV neutrinos which is unwanted. For harder spectral index, one needs to assume larger values of $\langle z_s \rangle$ to be compatible. The attenuated flux is shown in Fig. \ref{res}. 

\section{Conclusion}

To reconcile a light sterile neutrino of the type observed by MiniBooNE with BBN predictions, one must introduce gauge or scalar mediated interactions between the sterile neutrinos. Because of the lightness of the mediators required, there will be observable effects in the spectrum of high-energy neutrinos detected by IceCube. We have shown that the gaps in the spectrum at 400-800 TeV as well as beyond 2.6 PeV correspond to resonant absorption of two heaviest mass eigenstates. The prediction for the model at IceCube are peaks beyond 6.3 PeV and dips corresponding to two lighter neutrino mass states. These features may be observable in future IceCube data. A generic feature of absorption during propagation is that energy gap in the spectrum widens with distance to the source. This renders IceCube invisible to $\nu$ sources beyond a certain $z_{max}$. Future multi-messenger observations should be able to confirm this.

\end{document}